\documentclass{article}

    \PassOptionsToPackage{numbers, compress}{natbib}

\usepackage[preprint]{neurips_2024}

\usepackage[utf8]{inputenc} 
\usepackage[T1]{fontenc}    
\usepackage{hyperref}       
\usepackage{url}            
\usepackage{booktabs}       
\usepackage{amsfonts}       
\usepackage{nicefrac}       
\usepackage{microtype}      
\usepackage{xcolor}         
\usepackage{graphicx}
\usepackage{amsmath}
\usepackage{multirow}
\usepackage{threeparttable}

\title{Tree-of-Code: A Hybrid Approach for Robust Complex Task Planning and Execution
}

\author{%
  Yifan Li$^*$\\
  Global Innovation Exchange Institution, Tsinghua University \\
  Beijing, China, 100084 \\
  \texttt{yifan-li23@mail.tsinghua.edu.cn} \\ 
  \And
Ziyi Ni\thanks{The first two authors contributed equally to this work.  $\dagger$~Corresponding author. \texttt{dongdaxiang@baidu.com}} \\
  Institute of Automation, Chinese Academy of Sciences\\
  Beijing, China, 100190 \\
  \texttt{niziyi2021@ia.ac.cn} \\
  \AND
  Daxiang Dong$^{\dagger}$\\
  Baidu, Inc. \\
  Beijing, China, 100193‌ \\
  \texttt{dongdaxiang@baidu.com} \\
}

\begin{document}

\maketitle

\begin{abstract}
    The exceptional capabilities of large language models (LLMs) have substantially accelerated the rapid rise and widespread adoption of agents. Recent studies have demonstrated that generating Python code to consolidate LLM-based agents' actions into a unified action space (CodeAct) is a promising approach for developing real-world LLM agents. 
    However, this step-by-step code generation approach often lacks consistency and robustness, leading to instability in agent applications, particularly for complex reasoning and out-of-domain tasks. 
    In this paper, we propose a novel approach called Tree-of-Code (ToC) to tackle the challenges of complex problem planning and execution with an end-to-end mechanism. 
    By integrating key ideas from both Tree-of-Thought and CodeAct, ToC combines their strengths to enhance solution exploration. In our framework, each final code execution result is treated as a node in the decision tree, with a breadth-first search strategy employed to explore potential solutions.
    The final outcome is determined through a voting mechanism based on the outputs of the nodes.
    Experimental results on complicated task datasets demonstrate that our method provides more stable results compared to Tree-of-Thought and achieves higher accuracy than CodeAct. 
\end{abstract}

\section{Introduction}
In recent years, the application of code generation techniques to complex task planning and execution has garnered significant attention~\cite{holtl2mac,wen2024learning,xu2024wizardlm}, particularly with the emergence of CodeAct~\cite{wang2024codeact} approaches. CodeAct has demonstrated remarkable efficiency in generating executable code for complex tasks, improving overall performance in terms of speed and accuracy. 
However, the lack of consistent reasoning in CodeAct leads to frequent interruptions during multi-step generation, causing fragmented and stalled thinking. Additionally, this process accumulates significant model hallucinations~\cite{ji2023survey}, with increasing randomness, which ultimately undermines the robustness required for reliably solving complex problems. 

To address this challenge, we developed an end-to-end thought-code-execution pipeline that enables models to autonomously generate plans and decompositions for complex tasks, with clear reasoning. Post planning, the model produces implementation code that references earlier reasoning and leverages its innate code processing capabilities, making implicit reasoning explicit through code. Execution of this code yields results. Inspired by the Tree-of-Thought paradigm, our approach emphasizes structured solution exploration using decision trees and incorporates continuous code reflection. The main approach can be summarized into three parts:
\begin{enumerate}
\item End-to-End Code Generation: By producing complete code solutions end-to-end, we minimize the need for intermediate reflection on execution results, thus enhancing stability, and design a long thought-code reasoning process. We introduce `llm-function` to enable large models to generate prompts and summarize corresponding outcomes.
   
\item Exploration of Incomplete Nodes: Leveraging the Tree-of-Thought methodology, we explore nodes with incomplete execution results by varying prompts, large language models, and model temperatures to improve result stability.
   
\item Majority Voting for Final Results: large language models perform majority voting on all successfully executed nodes to determine the outcome.
\end{enumerate}

The contribution of this paper: 1) We propose the Tree-of-Code method, enhancing stability in complex task execution through efficient model integration. 2) Our approach can integrate various large language models without the need to do fine-tuning. 3) empirical validation of improved problem-solving performance.

\section{Related Work}
\textbf{LLM Reasoning:} 
Enhancing reasoning capabilities is a crucial step toward improving general LLMs. The current mainstream approaches focus on training techniques—such as pretraining \cite{han2020pretrain}, SFT \cite{chung2024flant5}, prompt-tuning \cite{han2020pretrain}, model editing \cite{thum2009editing}, and reinforcement learning \cite{snell2024rlscaling, zelikman2024quiet}—as well as the widely used method of prompt engineering to reduce output hallucinations and improve reasoning \cite{chen2023prompteng}. The latter primarily involves designing various frameworks for structured chains of thought. Longer thought chains seem to always yield better results \cite{zelikman2024quiet, wei2022cot}. The recent success of GPT-o1 has also highlighted the importance of extended chains of thought (CoT) \cite{wei2022cot}. Notably, existing CoT method, even with simple prompts like "let's think step by step," can achieve over a 30\% improvement. The Tree of Thoughts (ToT) framework \cite{yao2024tree} further enhances LLM reasoning by generating and evaluating intermediate steps, employing search algorithms to systematically explore and backtrack thought paths, enabling more effective problem-solving. Other emerging frameworks, like Graph-of-Thought \cite{besta2024got} and Algorithm-of-Thought \cite{sel2023aot}, are also contributing to this area of research.
\textbf{Code Generation: } 
It is widely accepted that training LLMs with codes can significantly enhance their reasoning abilities~\cite{chung2024flant5,achiam2023gpt,glm2024chatglm,guo2024deepseek,zhu2024deepseek}. Furthermore, research has shown that code, as a formal programming language, closely mirrors logical structures, inherently containing flow control mechanisms that can stimulate reasoning capabilities. For instance, \cite{wang2024codeact,guo2024deepseek, wang2023leti} argue that code serves as an effective medium for representing reasoning processes. Recent works combining code with agents have mainly focused on task completion in purely programming-related domains, such as software development \cite{wang2023leti,qian2024chatdev} and assisting humans in competitive programming tasks \cite{wen2024program,islam2024mapcoder}. 
However, these models do not use code as a scalable language, such as JSON, for direct communication in everyday tasks. CodeAct~\cite{wang2024codeact} attempted to use code directly to solve entire tasks, but the granularity of each step was too small, leading to error accumulation during function calls.

\begin{figure}
  \centering
  \includegraphics[width=0.8\textwidth]{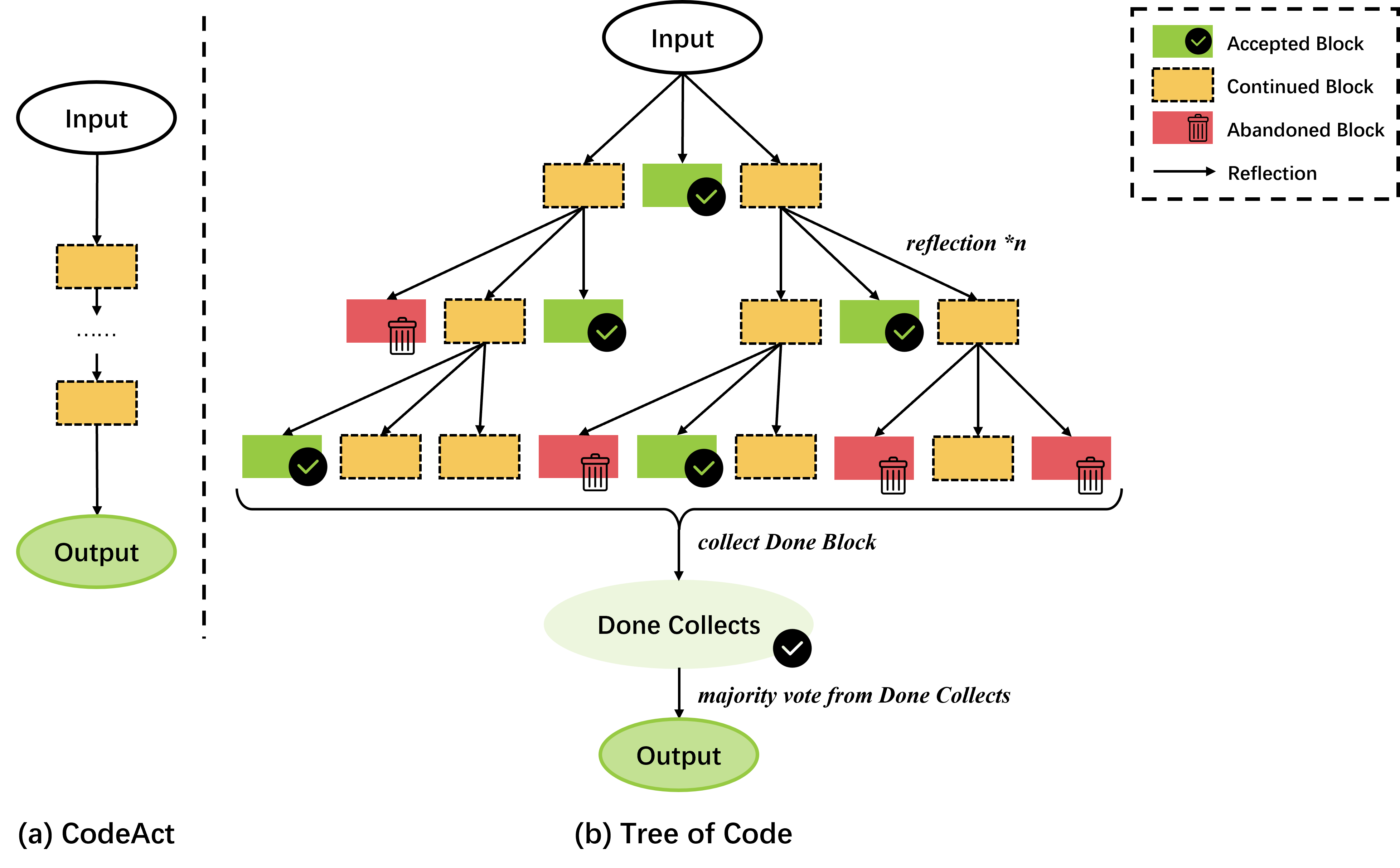}
  \caption{An Overview of our method \textbf{ToC} and \textbf{CodAct} comparisons.
(a) CodeAct receives input and performs a cycle of execution and correction, but the process is carried out in an iterative, round-by-round manner.
(b) ToC applies execution-level reflection in the decision-tree structure. At each layer, different nodes are executed in parallel; if executed correctly, they are stored in the candidate pool for voting,
if a node fails, it requires further reflection.
Yellow blocks mean continued reflection. Both red and green blocks are done: red blocks are discarded by LLM voting, while green blocks are collected and accepted. }
\end{figure}

\section{Tree-of-Code: Code Generation Through Tree-Structured Reflection}
\label{headings}

The Tree-of-Code (ToC) method combines the structured exploration of Tree-of-Thought with the efficient task planning of CodeAct. ToC introduces a tree-based generation and exploration mechanism that utilizes various code generation strategies and models, enhancing the robustness of execution results. By treating code as a form of reasoning, ToC capitalizes on the unique characteristics of code, such as consistency and determinism, to develop a more effective agent system with improved interpretability and reduced AI hallucinations.
This approach acknowledges that well-structured planning does not inherently ensure correct code execution. Unlike traditional reasoning methods, which rely on pre-organized plans, code necessitates a distinct methodology. Therefore, the code-reasoning process is decomposed into a structured framework that emphasizes reflection from post-execution feedback to inform iterative improvements. However, this direct treatment of code as logical reasoning introduces two significant challenges.
1) Limited diversity: Code generation tends to follow a single path, lacking exploration of alternative branches due to the inertia of models.
2) Limited strategy: Code generation lacks a systematic approach to evaluate various options or apply feedback to refine them.

To address these shortcomings, we introduce the Tree of Code (ToC), a framework that allows models to explore multiple solution paths using a variety of strategies and models, with reflection on post-generation execution built into the process.

\subsection{Overview the Tree-of-Code System}

The Tree-of-Code (ToC) framework presents a comprehensive code reasoning process comprising four key stages: thought, code generation, execution, and reflection. We represent this reasoning process as a tree $ T = (N, L) $
In contrast to the thought chain series, where thought steps are organized into nodes ($N$), our approach emphasizes end-to-end generation. Here, we define nodes as the combination of code generation and its corresponding execution. This tight coupling between code and execution results ensures logical consistency and correctness while facilitating high-quality outcomes for subsequent model training.
To enhance post-generation reflection through iterative refinement, we model the reflection process as lines ($L$), which can encompass various strategies for improvement: 1) System-level reflection: Sampling from diverse models to explore multiple solutions. 2) Operation-level reflection: Modifying evaluation and reflection strategies to iterative enhance output.
This structured approach guarantees that the final output is both accurate and diverse, striking a balance between correctness and a variety of reasoning pathways to yield robust solutions.

\subsection{Thought And Code Generator}
Code generation is a critical component of code-as-reasoning. In our framework, the thought and code generation stage integrates interactions between the user, environment, and agent: The user provides the task queries. The environment executes the generated code, providing execution feedback. The agent synthesizes information from the user, environment, and history, translating this into codes, which it then executes.
The process of transforming thought into executable codes can be formalized as:
\begin{equation}
\text{Execution}(i) = \text{Code}(\text{FC}(i), \text{FR}(i), \text{Thought}(i), \text{Functions}) \quad \forall i \in \{1, \dots, n\}
\end{equation}
where $Thought(i)$ represents the agent’s internal reasoning at step $i$, and $Code$ ($FC(i)$, $FR(i) $ and $Thought(i)$, $Functions$) converts that reasoning into Python code for execution. $FC(i)$ represents the code of father node of the current node. $FR(i)$ represents the code execution result of father node of the current node. $Functions$ represent all the tools that current task can use. 

To generate an end-to-end solution for current task, we add a llm-function tool into Functions set. With llm-function, our code generation LLMs will generate prompt from the current context and call llm-function to generate final results.

\subsection{Tree Expansion and State Evaluation}
\vspace{-2pt}
We initialize from a root node and do tree expansion recursively based on termination criteria. Every time a node's code is executed, state evaluation will be executed based on the code execution results. The process is as follows:
1) The process starts at the root node, expanding the tree incrementally, with each node representing generated thoughts, code, and execution results.
2) Generated codes of each leaf node will be executed in a Python environment. When errors are detected, child nodes are expanded from the erroneous node to analyze specific issues and explore alternative solutions.
3) This expansion continues until all leaf nodes execute successfully or a maximum depth is reached.
This strategy effectively manages complexity and ensures correct and efficient code execution results.

\subsection{Final Result Generator}
\vspace{-2pt}
After candidate outputs from all executed nodes are collected, these results undergo a majority vote and tragic summarization process to determine the most likely single answer. 
Explanations of the final answer are not included, just the precise answer to the question is shown.

\section{Experiment and Analysis}
\vspace{-4pt}
Our experiments evaluated the effectiveness of the ToC framework, comparing its performance primarily against the CodeAct framework. We used the M\textsuperscript{3}ToolEval, a newly curated benchmark for evaluating performance across complex multi-scene tasks, the same used in CodeAct~\cite{wang2024codeact}, to benchmark ToC. We evaluated the input samples from M\textsuperscript{3}ToolEval. The context window was fixed at 3k tokens, and the generation depth was set at 3 for consistency in computational costs. 
To highlight the superiority of our method, we directly compared it to the best-performing model in CodeAct, which is based on GPT-4. 
In the tests, ToC achieved a 7.2\% higher accuracy than CodeAct in tasks, and it reduced interaction steps significantly, demonstrating overall effectiveness and robustness in task completion, as Table \ref{tab:model_performance} shows.

To ensure diverse outputs, we applied a multi-level sampling strategy:
1) Model Diversity: A range of generative models (e.g., GPT-4~\cite{gpt4o}, ERNIE-4.0-Turbo~\cite{eb4t}, DeepSeek Coder~\cite{deepseek}, and Claude 3.5~\cite{claude35}) was utilized to introduce variability at the model level. 
2) Temperature Variability: Adjustments to generation temperatures were made within specified bounds to encourage content variation. 
3) Prompt Variation: Different role instructions and adaptive strategies were explored to further enhance adaptive diversity. 

Further analysis compared ToC against both the JSON-based and thought-based modes of CodeAct. Across all criteria, ToC consistently outperformed other traditional reasoning frameworks, particularly in managing complex reasoning tasks.

\vspace{-2pt}
\begin{table}[h]
  \caption{Performance Comparison}
  \label{tab:model_performance}
  \centering
  \begin{threeparttable}
  \begin{tabular}{llll}
    \toprule
    Model Name & Action Mode & Average Turns & Correct \% \\
    \midrule
    ToC\tnote{1} & code\_as\_reasoning & 2.3 & 81.60\% \\
    \midrule
    \multirow{3}{*}{CodeAct\tnote{2}} & code\_as\_reasoning & 5.5 & 74.40\% \\
    & json\_as\_reasoning & 7.6 & 52.40\% \\
    & text\_as\_reasoning & 7.7 & 53.70\% \\
    \bottomrule
  \end{tabular}
  \begin{tablenotes}
  \scriptsize 
    \item[1] Mix-modal sampling for Toc.
    \item[2] GPT-4, best performance model for CodeAct.
  \end{tablenotes}
  \end{threeparttable}
\end{table}
\vspace{-12pt}

\section{Conclusion}
\vspace{-4pt}
In this paper, we introduced the Tree-of-Code (ToC) method, which combines the strengths of Tree-of-Thought and CodeAct to enhance model robustness and accuracy. Both thought and code generation in our approach might intrigue internal processes of reasoning, while the decision-tree structured exploration enables reflection based on external execution. With efficient model integration and prompt optimization, we achieved results significantly surpassing the baselines on complex task datasets. Furthermore, our ongoing, yet unpublished, work suggests that this approach will yield similarly promising results in real-world applications and better performance after few-shot SFT. These results will be shared in future work—stay tuned.

{
\small

\bibliographystyle{plainnat}
\bibliography{neurips_2024}

\begin{thebibliography}{27}
\providecommand{\natexlab}[1]{#1}
\providecommand{\url}[1]{\texttt{#1}}
\expandafter\ifx\csname urlstyle\endcsname\relax
  \providecommand{\doi}[1]{doi: #1}\else
  \providecommand{\doi}{doi: \begingroup \urlstyle{rm}\Url}\fi

\bibitem[Achiam et~al.(2023{\natexlab{a}})Achiam, Adler, Agarwal, Ahmad, Akkaya, Aleman, Almeida, Altenschmidt, Altman, Anadkat, et~al.]{achiam2023gpt}
Josh Achiam, Steven Adler, Sandhini Agarwal, Lama Ahmad, Ilge Akkaya, Florencia~Leoni Aleman, Diogo Almeida, Janko Altenschmidt, Sam Altman, Shyamal Anadkat, et~al.
\newblock Gpt-4 technical report.
\newblock \emph{arXiv preprint arXiv:2303.08774}, 2023{\natexlab{a}}.

\bibitem[Achiam et~al.(2023{\natexlab{b}})Achiam, Adler, Agarwal, Ahmad, Akkaya, Aleman, Almeida, Altenschmidt, Altman, Anadkat, et~al.]{gpt4o}
Josh Achiam, Steven Adler, Sandhini Agarwal, Lama Ahmad, Ilge Akkaya, Florencia~Leoni Aleman, Diogo Almeida, Janko Altenschmidt, Sam Altman, Shyamal Anadkat, et~al.
\newblock Gpt-4 technical report.
\newblock \emph{arXiv preprint arXiv:2303.08774}, 2023{\natexlab{b}}.

\bibitem[Anthropic(2023)]{claude35}
Anthropic.
\newblock Claude 3.5, 2023.
\newblock URL \url{https://www.anthropic.com}.

\bibitem[Besta et~al.(2024)Besta, Blach, Kubicek, Gerstenberger, Podstawski, Gianinazzi, Gajda, Lehmann, Niewiadomski, Nyczyk, et~al.]{besta2024got}
Maciej Besta, Nils Blach, Ales Kubicek, Robert Gerstenberger, Michal Podstawski, Lukas Gianinazzi, Joanna Gajda, Tomasz Lehmann, Hubert Niewiadomski, Piotr Nyczyk, et~al.
\newblock Graph of thoughts: Solving elaborate problems with large language models.
\newblock In \emph{Proceedings of the AAAI Conference on Artificial Intelligence}, volume~38, pages 17682--17690, 2024.

\bibitem[Chen et~al.(2023)Chen, Zhang, Langren{\'e}, and Zhu]{chen2023prompteng}
Banghao Chen, Zhaofeng Zhang, Nicolas Langren{\'e}, and Shengxin Zhu.
\newblock Unleashing the potential of prompt engineering in large language models: a comprehensive review.
\newblock \emph{arXiv preprint arXiv:2310.14735}, 2023.

\bibitem[Chung et~al.(2024)Chung, Hou, Longpre, Zoph, Tay, Fedus, Li, Wang, Dehghani, Brahma, et~al.]{chung2024flant5}
Hyung~Won Chung, Le~Hou, Shayne Longpre, Barret Zoph, Yi~Tay, William Fedus, Yunxuan Li, Xuezhi Wang, Mostafa Dehghani, Siddhartha Brahma, et~al.
\newblock Scaling instruction-finetuned language models.
\newblock \emph{Journal of Machine Learning Research}, 25\penalty0 (70):\penalty0 1--53, 2024.

\bibitem[GLM et~al.(2024)GLM, Zeng, Xu, Wang, Zhang, Yin, Rojas, Feng, Zhao, Lai, et~al.]{glm2024chatglm}
Team GLM, Aohan Zeng, Bin Xu, Bowen Wang, Chenhui Zhang, Da~Yin, Diego Rojas, Guanyu Feng, Hanlin Zhao, Hanyu Lai, et~al.
\newblock Chatglm: A family of large language models from glm-130b to glm-4 all tools.
\newblock \emph{arXiv preprint arXiv:2406.12793}, 2024.

\bibitem[Guo et~al.(2024{\natexlab{a}})Guo, Zhu, Yang, Xie, Dong, Zhang, Chen, Bi, Wu, Li, et~al.]{deepseek}
Daya Guo, Qihao Zhu, Dejian Yang, Zhenda Xie, Kai Dong, Wentao Zhang, Guanting Chen, Xiao Bi, Yu~Wu, YK~Li, et~al.
\newblock Deepseek-coder: When the large language model meets programming--the rise of code intelligence.
\newblock \emph{arXiv preprint arXiv:2401.14196}, 2024{\natexlab{a}}.

\bibitem[Guo et~al.(2024{\natexlab{b}})Guo, Zhu, Yang, Xie, Dong, Zhang, Chen, Bi, Wu, Li, et~al.]{guo2024deepseek}
Daya Guo, Qihao Zhu, Dejian Yang, Zhenda Xie, Kai Dong, Wentao Zhang, Guanting Chen, Xiao Bi, Yu~Wu, YK~Li, et~al.
\newblock Deepseek-coder: When the large language model meets programming--the rise of code intelligence.
\newblock \emph{arXiv preprint arXiv:2401.14196}, 2024{\natexlab{b}}.

\bibitem[Han et~al.(2020)Han, Ren, and Peng]{han2020pretrain}
Rujun Han, Xiang Ren, and Nanyun Peng.
\newblock Econet: Effective continual pretraining of language models for event temporal reasoning.
\newblock \emph{arXiv preprint arXiv:2012.15283}, 2020.

\bibitem[Holt et~al.(2024)Holt, Luyten, and van~der Schaar]{holtl2mac}
Samuel Holt, Max~Ruiz Luyten, and Mihaela van~der Schaar.
\newblock L2mac: Large language model automatic computer for extensive code generation.
\newblock In \emph{The Twelfth International Conference on Learning Representations}, 2024.

\bibitem[Islam et~al.(2024)Islam, Ali, and Parvez]{islam2024mapcoder}
Md~Ashraful Islam, Mohammed~Eunus Ali, and Md~Rizwan Parvez.
\newblock Mapcoder: Multi-agent code generation for competitive problem solving.
\newblock \emph{arXiv preprint arXiv:2405.11403}, 2024.

\bibitem[Ji et~al.(2023)Ji, Lee, Frieske, Yu, Su, Xu, Ishii, Bang, Madotto, and Fung]{ji2023survey}
Ziwei Ji, Nayeon Lee, Rita Frieske, Tiezheng Yu, Dan Su, Yan Xu, Etsuko Ishii, Ye~Jin Bang, Andrea Madotto, and Pascale Fung.
\newblock Survey of hallucination in natural language generation.
\newblock \emph{ACM Computing Surveys}, 55\penalty0 (12):\penalty0 1--38, 2023.

\bibitem[Qian et~al.(2024)Qian, Liu, Liu, Chen, Dang, Li, Yang, Chen, Su, Cong, et~al.]{qian2024chatdev}
Chen Qian, Wei Liu, Hongzhang Liu, Nuo Chen, Yufan Dang, Jiahao Li, Cheng Yang, Weize Chen, Yusheng Su, Xin Cong, et~al.
\newblock Chatdev: Communicative agents for software development.
\newblock In \emph{Proceedings of the 62nd Annual Meeting of the Association for Computational Linguistics (Volume 1: Long Papers)}, pages 15174--15186, 2024.

\bibitem[Ren et~al.(2023)Ren, Li, and Duan]{eb4t}
Chengxiang Ren, Yingbo Li, and Yucong Duan.
\newblock Evaluation on agi/gpt based on the dikwp for ernie bot.
\newblock \emph{arXiv preprint}, 2023.

\bibitem[Sel et~al.(2023)Sel, Al-Tawaha, Khattar, Jia, and Jin]{sel2023aot}
Bilgehan Sel, Ahmad Al-Tawaha, Vanshaj Khattar, Ruoxi Jia, and Ming Jin.
\newblock Algorithm of thoughts: Enhancing exploration of ideas in large language models.
\newblock \emph{arXiv preprint arXiv:2308.10379}, 2023.

\bibitem[Snell et~al.(2024)Snell, Lee, Xu, and Kumar]{snell2024rlscaling}
Charlie Snell, Jaehoon Lee, Kelvin Xu, and Aviral Kumar.
\newblock Scaling llm test-time compute optimally can be more effective than scaling model parameters.
\newblock \emph{arXiv preprint arXiv:2408.03314}, 2024.

\bibitem[Thum et~al.(2009)Thum, Batory, and Kastner]{thum2009editing}
Thomas Thum, Don Batory, and Christian Kastner.
\newblock Reasoning about edits to feature models.
\newblock In \emph{2009 IEEE 31st International Conference on Software Engineering}, pages 254--264. IEEE, 2009.

\bibitem[Wang et~al.(2023)Wang, Peng, Jabbarvand, and Ji]{wang2023leti}
Xingyao Wang, Hao Peng, Reyhaneh Jabbarvand, and Heng Ji.
\newblock Leti: Learning to generate from textual interactions.
\newblock \emph{arXiv preprint arXiv:2305.10314}, 2023.

\bibitem[Wang et~al.(2024)Wang, Chen, Yuan, Zhang, Li, Peng, and Ji]{wang2024codeact}
Xingyao Wang, Yangyi Chen, Lifan Yuan, Yizhe Zhang, Yunzhu Li, Hao Peng, and Heng Ji.
\newblock Executable code actions elicit better llm agents.
\newblock \emph{arXiv preprint arXiv:2402.01030}, 2024.

\bibitem[Wei et~al.(2022)Wei, Wang, Schuurmans, Bosma, Xia, Chi, Le, Zhou, et~al.]{wei2022cot}
Jason Wei, Xuezhi Wang, Dale Schuurmans, Maarten Bosma, Fei Xia, Ed~Chi, Quoc~V Le, Denny Zhou, et~al.
\newblock Chain-of-thought prompting elicits reasoning in large language models.
\newblock \emph{Advances in neural information processing systems}, 35:\penalty0 24824--24837, 2022.

\bibitem[Wen et~al.(2024{\natexlab{a}})Wen, Zhong, Ke, Shao, Wang, and Huang]{wen2024learning}
Jiaxin Wen, Ruiqi Zhong, Pei Ke, Zhihong Shao, Hongning Wang, and Minlie Huang.
\newblock Learning task decomposition to assist humans in competitive programming.
\newblock In \emph{Proceedings of the 62nd Annual Meeting of the Association for Computational Linguistics}, 2024{\natexlab{a}}.

\bibitem[Wen et~al.(2024{\natexlab{b}})Wen, Zhong, Ke, Shao, Wang, and Huang]{wen2024program}
Jiaxin Wen, Ruiqi Zhong, Pei Ke, Zhihong Shao, Hongning Wang, and Minlie Huang.
\newblock Learning task decomposition to assist humans in competitive programming.
\newblock \emph{arXiv preprint arXiv:2406.04604}, 2024{\natexlab{b}}.

\bibitem[Xu et~al.(2024)Xu, Sun, Zheng, Geng, Zhao, Feng, Tao, Lin, and Jiang]{xu2024wizardlm}
Can Xu, Qingfeng Sun, Kai Zheng, Xiubo Geng, Pu~Zhao, Jiazhan Feng, Chongyang Tao, Qingwei Lin, and Daxin Jiang.
\newblock Wizardlm: Empowering large pre-trained language models to follow complex instructions.
\newblock In \emph{The Twelfth International Conference on Learning Representations}, 2024.

\bibitem[Yao et~al.(2024)Yao, Yu, Zhao, Shafran, Griffiths, Cao, and Narasimhan]{yao2024tree}
Shunyu Yao, Dian Yu, Jeffrey Zhao, Izhak Shafran, Tom Griffiths, Yuan Cao, and Karthik Narasimhan.
\newblock Tree of thoughts: Deliberate problem solving with large language models.
\newblock \emph{Advances in Neural Information Processing Systems}, 36, 2024.

\bibitem[Zelikman et~al.(2024)Zelikman, Harik, Shao, Jayasiri, Haber, and Goodman]{zelikman2024quiet}
Eric Zelikman, Georges Harik, Yijia Shao, Varuna Jayasiri, Nick Haber, and Noah~D Goodman.
\newblock Quiet-star: Language models can teach themselves to think before speaking.
\newblock \emph{arXiv preprint arXiv:2403.09629}, 2024.

\bibitem[Zhu et~al.(2024)Zhu, Guo, Shao, Yang, Wang, Xu, Wu, Li, Gao, Ma, et~al.]{zhu2024deepseek}
Qihao Zhu, Daya Guo, Zhihong Shao, Dejian Yang, Peiyi Wang, Runxin Xu, Y~Wu, Yukun Li, Huazuo Gao, Shirong Ma, et~al.
\newblock Deepseek-coder-v2: Breaking the barrier of closed-source models in code intelligence.
\newblock \emph{arXiv preprint arXiv:2406.11931}, 2024.

\end{thebibliography}

}






\newpage
\section*{NeurIPS Paper Checklist}

The checklist is designed to encourage best practices for responsible machine learning research, addressing issues of reproducibility, transparency, research ethics, and societal impact. Do not remove the checklist: {\bf The papers not including the checklist will be desk rejected.} The checklist should follow the references and follow the (optional) supplemental material.  The checklist does NOT count towards the page
limit. 

Please read the checklist guidelines carefully for information on how to answer these questions. For each question in the checklist:
\begin{itemize}
    \item You should answer \answerYes{}, \answerNo{}, or \answerNA{}.
    \item \answerNA{} means either that the question is Not Applicable for that particular paper or the relevant information is Not Available.
    \item Please provide a short (1–2 sentence) justification right after your answer (even for NA). 
\end{itemize}

{\bf The checklist answers are an integral part of your paper submission.} They are visible to the reviewers, area chairs, senior area chairs, and ethics reviewers. You will be asked to also include it (after eventual revisions) with the final version of your paper, and its final version will be published with the paper.

The reviewers of your paper will be asked to use the checklist as one of the factors in their evaluation. While "\answerYes{}" is generally preferable to "\answerNo{}", it is perfectly acceptable to answer "\answerNo{}" provided a proper justification is given (e.g., "error bars are not reported because it would be too computationally expensive" or "we were unable to find the license for the dataset we used"). In general, answering "\answerNo{}" or "\answerNA{}" is not grounds for rejection. While the questions are phrased in a binary way, we acknowledge that the true answer is often more nuanced, so please just use your best judgment and write a justification to elaborate. All supporting evidence can appear either in the main paper or the supplemental material, provided in appendix. If you answer \answerYes{} to a question, in the justification please point to the section(s) where related material for the question can be found.

IMPORTANT, please:
\begin{itemize}
    \item {\bf Delete this instruction block, but keep the section heading ``NeurIPS paper checklist"},
    \item  {\bf Keep the checklist subsection headings, questions/answers and guidelines below.}
    \item {\bf Do not modify the questions and only use the provided macros for your answers}.
\end{itemize}


\begin{enumerate}

\item {\bf Claims}
    \item[] Question: Do the main claims made in the abstract and introduction accurately reflect the paper's contributions and scope?
    \item[] Answer: \answerYes{} 
    \item[] Justification: The paper claims main contributions and scope in the abstract and introduction.

\item {\bf Limitations}
    \item[] Question: Does the paper discuss the limitations of the work performed by the authors?
    \item[] Answer: \answerYes{} 
    \item[] Justification: Yes, the paper discusses the limitations of the work.

\item {\bf Theory Assumptions and Proofs}
    \item[] Question: For each theoretical result, does the paper provide the full set of assumptions and a complete (and correct) proof?
    \item[] Answer: \answerNA{} 
    \item[] Justification: Our paper mainly works on method and experiments and do not include complete proof about the method.

    \item {\bf Experimental Result Reproducibility}
    \item[] Question: Does the paper fully disclose all the information needed to reproduce the main experimental results of the paper to the extent that it affects the main claims and/or conclusions of the paper (regardless of whether the code and data are provided or not)?
    \item[] Answer: \answerYes{} 
    \item[] Justification: the paper will open source code on github for other researchers to reproduce experiments.

\item {\bf Open access to data and code}
    \item[] Question: Does the paper provide open access to the data and code, with sufficient instructions to faithfully reproduce the main experimental results, as described in supplemental material?
    \item[] Answer: \answerYes{} 
    \item[] Justification: the paper will provide the open source code.

\item {\bf Experimental Setting/Details}
    \item[] Question: Does the paper specify all the training and test details (e.g., data splits, hyperparameters, how they were chosen, type of optimizer, etc.) necessary to understand the results?
    \item[] Answer: \answerYes{} 
    \item[] Justification: Yes, the paper used open sourced dataset and include all the details of the datasets.

\item {\bf Experiment Statistical Significance}
    \item[] Question: Does the paper report error bars suitably and correctly defined or other appropriate information about the statistical significance of the experiments?
    \item[] Answer: \answerYes{} 
    \item[] Justification: Yes, the paper reported error bars properly.

\item {\bf Experiments Compute Resources}
    \item[] Question: For each experiment, does the paper provide sufficient information on the computer resources (type of compute workers, memory, time of execution) needed to reproduce the experiments?
    \item[] Answer: \answerYes{} 
    \item[] Justification: Yes, the paper includes all the computational resources information.
    
\item {\bf Code Of Ethics}
    \item[] Question: Does the research conducted in the paper conform, in every respect, with the NeurIPS Code of Ethics \url{https://neurips.cc/public/EthicsGuidelines}?
    \item[] Answer: \answerYes{} 
    \item[] Justification: Yes, the research conducted in the paper conform, in every respect, with the NeurIPS Code of Ethics.

\item {\bf Broader Impacts}
    \item[] Question: Does the paper discuss both potential positive societal impacts and negative societal impacts of the work performed?
    \item[] Answer: \answerYes{} 
    \item[] Justification: Yes, the paper discussed the issue.
    
\item {\bf Safeguards}
    \item[] Question: Does the paper describe safeguards that have been put in place for responsible release of data or models that have a high risk for misuse (e.g., pretrained language models, image generators, or scraped datasets)?
    \item[] Answer: \answerNA{} 
    \item[] Justification: We mainly use open sourced models and commercialized model API. These models commonly have safeguards in their own license description. 

\item {\bf Licenses for existing assets}
    \item[] Question: Are the creators or original owners of assets (e.g., code, data, models), used in the paper, properly credited and are the license and terms of use explicitly mentioned and properly respected?
    \item[] Answer: \answerYes{} 
    \item[] Justification: Yes, our open sourced code will be licensed properly.

\item {\bf New Assets}
    \item[] Question: Are new assets introduced in the paper well documented and is the documentation provided alongside the assets?
    \item[] Answer: \answerYes{} 
    \item[] Justification: We introduced a new method for the problem and will release our code on github if accepted.

\item {\bf Crowdsourcing and Research with Human Subjects}
    \item[] Question: For crowdsourcing experiments and research with human subjects, does the paper include the full text of instructions given to participants and screenshots, if applicable, as well as details about compensation (if any)? 
    \item[] Answer: \answerNA{} 
    \item[] Justification: Our paper does experiment on existing open sourced dataset and the evaluation code of the method is on github.

\item {\bf Institutional Review Board (IRB) Approvals or Equivalent for Research with Human Subjects}
    \item[] Question: Does the paper describe potential risks incurred by study participants, whether such risks were disclosed to the subjects, and whether Institutional Review Board (IRB) approvals (or an equivalent approval/review based on the requirements of your country or institution) were obtained?
    \item[] Answer: \answerNA{} 
    \item[] Justification: We did not crowsourcing our research experiments.

\end{enumerate}

\end{document}